\documentclass[sigconf]{acmart}

\usepackage{graphicx}
\usepackage{todonotes}

\copyrightyear{2020}
\acmYear{2020}
\setcopyright{acmlicensed}
\acmConference[GECCO '20]{Genetic and Evolutionary Computation Conference}{July 8--12, 2020}{Cancún, Mexico}
\acmBooktitle{Genetic and Evolutionary Computation Conference (GECCO '20), July 8--12, 2020, Cancún, Mexico}
\acmPrice{15.00}
\acmDOI{10.1145/3377930.3390239}
\acmISBN{978-1-4503-7128-5/20/07}

\begin{document}
\title{Code Building Genetic Programming}

%%% The submitted version for review should be ANONYMOUS
\author{Edward Pantridge}
\orcid{0000-0003-0535-5268}
\affiliation{%
  \institution{Swoop}
  \city{Cambridge} 
  \state{Massachusetts} 
  \country{USA}
}
\email{ed@swoop.com}

\author{Lee Spector}
\orcid{0000-0001-5299-4797}
\affiliation{%
	\institution{Amherst College, Hampshire College, and UMass Amherst}
	\city{Amherst}
	\state{Massachusetts}
	\country{USA}
}
\email{lspector@hampshire.edu}

\begin{abstract}
In recent years the field of genetic programming  has made significant advances towards automatic programming. Research and development of contemporary program synthesis methods, such as PushGP and Grammar Guided Genetic Programming, can produce programs that solve problems typically assigned in introductory academic settings. These problems focus on a narrow, predetermined set of simple data structures,  basic control flow patterns, and primitive, non-overlapping data types (without, for example, inheritance or composite types). Few, if any, genetic programming methods for program synthesis have convincingly demonstrated the capability of synthesizing programs that use arbitrary data types, data structures, and specifications that are drawn from existing codebases. In this paper, we introduce Code Building Genetic Programming (CBGP) as a framework within which this can be done, by leveraging programming language features such as reflection and first-class specifications. CBGP produces a computational graph that can be executed or translated into source code of a host language. To demonstrate the novel capabilities of CBGP, we present results on new benchmarks that use non-primitive, polymorphic data types as well as some standard program synthesis benchmarks.
\end{abstract}




\maketitle

\section{Introduction}
\label{sec:intro}

``Inductive program synthesis'' is the term used to describe the process of constructing an executable program from a set of input-output examples. A form of inductive program synthesis which has been gaining attention in recent years is ``general program synthesis,'' which specifically aims to produce programs that use similar constructs (data types, control flows, data structures) as human programmers. The goal of this research area is to eventually discover a process of automatic programming that is comparable to human skill~\cite{oneill:2019:AutomaticProgrammingOpen}.

Real world applications of a sufficiently sophisticated automatic programming system would have dramatic impact on software development. Organizations could potentially deploy automated systems that attempt to fix bugs or reduce the resource utilization of their software. In addition, there may be a class of problems for which the solution is difficult for humans to produce, but is easy for program synthesis methods to discover.

An ideal program synthesis framework must be capable of producing programs that utilize arbitrary data types, including types with arbitrary relationships, such as inheritance. The computations performed by this system must support generic manipulation of polymorphic types.

Humans rely on decomposition, abstraction, and re-use to create complex software applications. An ideal program synthesis framework must be able to utilize preexisting abstractions (human written and/or previously synthesized) in the programs it produces. Furthermore, the framework should not require additional configuration or modification to use problem-specific abstractions.

The output of an ideal program synthesis framework would be the same as the output of a human programmer. Typically this is source code that can be executed in any environment that supports the programming language. 

No known program synthesis methodology is capable of achieving the goals stated above. This paper presents Code Building Genetic Programming (CBGP), a general program synthesis system with comparable problem solving performance to PushGP and G3P~\footnote{An open source implementation of our prototype can be found at https://github.com/erp12/CodeBuildingGeneticProgramming-ProtoType}. CBGP also provides additional capabilities that are not present in other program synthesis systems. These include: use of generic functions, support for polymorphic types, automatic integration with existing codebases, and transcription of programs into human-readable source code of the host language. 

The rest of this paper is structured as follows. First, we discuss related work that informed our research. Second, we describe CBGP's program representation and the architecture of the search process. We then provide details on the benchmark problems used to evaluate CBGP and present early comparisons against other methods. We conclude with discussion of the implications of this work and suggestions for future research.

\section{Related Work}

Inductive program synthesis is a research topic that spans multiple fields of study. Evolutionary computation has produced multiple genetic programming frameworks designed to synthesize general programs. Some of the state-of-the-art program synthesis methods include PushGP and Grammar Guided Genetic Programming (G3P).

PushGP evolves programs in a Turing complete language, called Push. The Push language uses a stack-based execution model~\cite{Spector:2005:push3}. Push programs are nested sequences of instructions and literals. Literals are values that get directly pushed onto a particular stack based on their data type. Instructions are functions that pop values off the stacks, transform them, and push the result back onto the appropriate stack. Programs are run through an interpreter and the final state of the stacks is considered the output of the program. Push code can be pushed onto a dedicated stack that can be used by instructions to implement control flow patterns like iteration and conditionals.

It is not possible for PushGP to evolve programs that use arbitrary data types without the definition of additional instructions. This means a custom PushGP system would need to be built to take advantage of preexisting problem-specific abstractions. Furthermore, it is not clear how to build a PushGP system capable of using overlapping types. If $C_{sub}$ is a concrete subclass with a concrete super-class, $C_{super}$, which stack should an instruction that requires an input of type $C_{super}$ pop from to find its arguments? To our knowledge, no PushGP implementation has ever developed a successful strategy for dealing with this situation or any other form of polymorphism.

PushGP programs are nested sequences of tokens that can be processed by a Push interpreter. Thus, the programs are only as portable as the PushGP implementations that produces them. Re-implementation of Push programs into source code of a host-language is difficult because the stack-based execution does not map cleanly into common programming paradigms.

Grammar Guided Genetic Programming (G3P) is a more recently introduced method of inductive program synthesis that features a set of context-free Backus-Naur Form grammars~\cite{Forstenlechner:2017:G3P}. G3P uses one grammar per supported data type to achieve coverage over general programs. Using these grammars, a grammatical evolution (GE) framework can map variable length sequences of integers into programs~\cite{Ryan:1998:GE}.

The G3P methodology has the advantage of producing type-safe source code in the host-language. The code produced can contain variable assignments, control flow, and iteration. It has been previously discussed that extension of G3P systems to support additional data types is laborious. It requires the manual creation and integration of additional grammars~\cite{Forstenlechner:2017:G3P}. It has also been shown that the performance of G3P is sensitive to the exact implementation of the grammars~\cite{Forstenlechner:2018:G3P-extention}. Similar to PushGP, we do not know of any prior work that attempts to extend G3P to support polymorphism.

Other program synthesis methods that are not variants of genetic programming exist, such as: Excel Flash Fill~\cite{gulwani:2011:flashfill}, MagicHaskeller~\cite{Katayama:2008:MagicHaskeller}, TerpreT~\cite{Gaunt:2016:terpret}, and DeepCoder~\cite{Balog:2016:DeepCoder}. Although all of these systems are successful in their target problem domains, they are incapable of attempting the standard program synthesis benchmark problems due to a lack of support for crucial data types and control structures~\cite{Pantridge:2017:IPS-benchmark}. 

Code Building Genetic Programming aims to be a program synthesis methodology with comparable search performance to PushGP and G3P, while adding some additional capabilities that bring the field significantly closer to feasible real-world applications.

\section{Structure of CBGP Programs}

Programs produced by Code Building Genetic Programming are computational graphs. Each inner node of a graph is a function and each leaf node is either a constant value or a reference to one of the program's inputs. We use the general term "expression" as the name for a node in a computational graph. The computational graphs are directed and acyclic (DAG). A program produced by CBGP will be a DAG of expressions, called a "program DAG."

\subsection{Expressions}

Expressions are containers that encapsulate a particular computation or value. In addition, expressions hold an associated specification of the computation they encapsulate. For this work, the specifications are annotations denoting relevant data types associated with the expression. 
% By encapsulating the computation with its specification into an expression, $E_1$, it can be known if the output of $E_1$ will be a valid input to some other expression, $E_2$. In other words, it can be confirmed that a valid program DAG, $G$, can be made with $E_2$ as the root and $E_1$ as the child of $E_2$.

The prototype implementation of CBGP discussed in this paper uses a finite set of expression types. These include: \textit{Constants, Inputs, Functions, Methods, Constructors}, and \textit{Higher Order Functions}. 

\begin{figure}
    \centering
    \includegraphics[width=0.85\columnwidth]{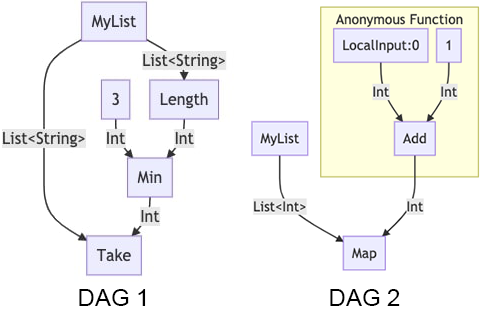}
    \caption{Some example program DAGS. ``DAG 1'' shows a program that will return the first 3 elements of a list of strings. The list of strings is provided to the DAG as an input value named \texttt{MyList}. Definitions for the expressions in ``DAG 1'' can be found in Figure~\ref{fig:first_three_expressions}. ``DAG 2'' is an example of the use of higher order functions described in Section \ref{sec:hof}.}
    \label{fig:dags}
\end{figure}

\begin{figure*}
    \begin{tabular}{lllll}
        Name & Expr. Type & Arguments & Return & Behavior \\ \hline
        MyList & Input & - & List[str] & An input that will be supplied each time the DAG is executed. \\
        3 & Constant & - & Int & A constant integer value of 3. \\
        Length & Method & $L$: List & Int & Returns the number of elements in $L$. \\
        Min & Function & $a$: Number, $b$: Number & Number & Returns the the minimum of $a$ and $b$. \\ 
        Take & Method & $L$: List, $N$: Int & List & Returns a list containing the first $N$ elements of $L$. \\
    \end{tabular}
    \caption{The definitions of the expressions found in ``DAG 1'' of Figure~\ref{fig:dags}. The ``Arguments'' and ``Return'' attributes of a an expression make up their specification. \textit{Method} expressions treat the class instance object as an implicit argument.}
    \label{fig:first_three_expressions}
\end{figure*}

DAG 1 shown in Figure~\ref{fig:dags} shows an example program DAG that represents a program that returns the first 3 elements of a list, or the entire list if it contains 3 or fewer elements. Descriptions of each expression in the DAG are found in Figure~\ref{fig:first_three_expressions}.

Note that the \texttt{MyList} expression is specified to return a list of strings which is a sub-type of list. The \texttt{Length} expression is specified to take a generic list as an argument, but the \texttt{MyList} expression is a valid child expression because all values of \texttt{MyList} are guaranteed to be instances of the expected argument type. Program DAGs are type-safe because the return type of all child expressions are sub-types (or the same type) as the corresponding argument types of their parent expression. 

\textit{Constants} and \textit{Inputs} are leaf nodes that don't require any arguments. \textit{Constants} always return the same value and \textit{Inputs} return the value of a specific input to the program DAG. Both expression types have a specification that is the data type of the value of the node.  If the data type is a collection type, union type, or some other polymorphic type, the specification will be a decompose-able representation of the type and its parts. For example, a key-value map type annotation can be decomposed into the data type of the keys and values respectively. In Figure \ref{fig:dags} the ``MyList'' expression is an \textit{Input} and the ``3'' expression is a \textit{Constant}.

Some expression nodes of a program DAG represent a function-like computation that accepts arguments and returns a value. These expressions wrap an underlying function in the host-language. \textit{Method} and \textit{Constructor} expressions wrap parts of a pre-existing class definition. All function-like expressions have the same structure in their specification consisting of of a mapping from argument names to types and an additional type corresponding to the return value. In Figure \ref{fig:dags}  ``Length,'' ``Min,'' ``Take,'' and ``Add'' are all function-like expressions.

It is possible for \textit{Function} and \textit{Constructor} expressions to be leaf nodes of a program DAG if they have an arity of zero. \textit{Method} expressions cannot be leaf nodes because they are treated as functions which have an implicit argument that is the class instance object.

\subsection{Higher Order and Anonymous Functions}
\label{sec:hof}

Contemporary program synthesis frameworks, such as PushGP and G3P, have demonstrated the capability to produce programs that utilize iteration and control flow. It is challenging to express generic forms of iteration, such as a while-loop, in an acyclic computational graph. In contrast, higher order functions are beginning to see more widespread use in the computational graphs used by big-data processing frameworks like Spark~\cite{databricks_spark_hof}. CBGP draws inspiration from these modern DAG representations and uses higher order functions in program DAGs to manipulate data structures.

% \begin{figure}
%     \centering
%     \includegraphics[width=0.45\columnwidth]{img/dag2_hof.png}
%     \caption{A program DAG that increments each element of a list of integers by 1. The nodes within the box represent the anonymous function body that is executed once per element of the collection.}
%     \label{fig:inc_each_dag}
% \end{figure}

Figure~\ref{fig:dags} depicts a program DAG that utilizes the \texttt{map} higher order function. The first argument to the \texttt{map} expression must be an expression that returns a collection. The second argument is an expression that represents an anonymous function to apply to each element of the collection. 

As shown in DAG 2 of Figure~\ref{fig:dags}, one of the expressions contained in the anonymous function sub-DAG is a special kind of \textit{Input} expression, called a ``local input.'' These expressions can only exist in a program DAG as part of an anonymous function. Each call to the anonymous function within an execution of the DAG will use the collection element as the value for the local input. It is possible for multiple local input expressions to appear in a single anonymous function sub-DAG, and anonymous function DAGs may contain higher order function expressions with their nested anonymous function DAGs.

When the \texttt{map} \textit{Higher Order Function} expression is evaluated as part of a program DAG, the following process is followed:

\begin{enumerate}
    \item An empty sequential buffer is initialized.
    \item The child expression corresponding to the sequential collection is evaluated.
    \item For each element of the resulting collection:
    \begin{enumerate}
        \item The anonymous function child expression is called with the element passed as the value for the local input. Input values of the overall program DAG are also passed to the anonymous function.
        \item The resulting value is appended to the buffer.
    \end{enumerate}
    \item The buffer is returned as the output of the \texttt{map} expression.
\end{enumerate}

The current implementation of CBGP also supports a \texttt{filter} \textit{Higher Order Function} expression which follows a similar process. The anonymous function for the \texttt{filter} expression must return a Boolean value. Additional \textit{Higher Order Function} expressions, such as reducing and partitioning, are possible but have not been implemented yet in any CBGP system.

\section{Automatic Expression Creation}
\label{sec:reflection}

It is becoming increasingly popular for programming languages to be accompanied by rich, first-class, data oriented, specification tooling. These tools enable static and run-time validations of code behavior. One example of this tooling is the \texttt{typing} module introduced into the standard library of the Python programming language as of version 3.5. This library adds type hint annotations to function definitions, which complement the primarily object oriented programming language~\cite{python_type_hints}. The Clojure \texttt{Spec} library is an example of an exceptionally expressive specification framework that goes beyond type checking~\cite{hickey_2016}. It has seen rapid adoption after its emergence into Clojure's functional programming ecosystem. These new technologies allow for systems to reason about programs, and their behavior, as first-class objects which is profoundly useful for program synthesis systems. 

The prototype implementation of Code Building Genetic Programming uses Python's type annotations as specifications to the values and functions encapsulated in expressions. Importantly, for polymorphic types the annotations can go beyond the generic data-type. For example, a list of strings might be annotated as \texttt{list} or as \texttt{List[str]}. The \texttt{List[str]} annotation can be decomposed as a collection type (\texttt{List}) and an element type (\texttt{str}). The \texttt{typing} system also supports the comparison of types at run-time. For example, \texttt{list} is a sub-type of \texttt{Sequence} by inheritance and \texttt{int} is a sub-type of \texttt{Union[int, float]} by composition.

Furthermore, language features such as reflection can be used to automatically discover the available functions, types, and values defined in the environment as well as their associated specifications. This allows CBGP to be injected into an arbitrary run-time environment, with existing code and external dependencies, and create a set of expressions from the discover-able classes, functions, and variables. The only assumption made by CBGP is that there is an available specification for everything it finds. The implication of this is that CBGP can synthesize programs which utilize human written code, previously synthesized code, and external libraries without any additional modification or configuration.

As a consequence, in real world applications the performance of CBGP can be improved by a human developer creating additional abstraction for the CBGP system to use. This kind of cooperation between a human programmer and a program synthesis framework is infeasible under any other known method because incorporation of new human written abstractions require modification or laborious configuration of the program synthesis method.

\section{Composition and Reification of Programs}

Once a set of expressions has been defined within the context of a Code Building Genetic Programming application, program DAGs can be produced via the composition of expressions. An expression becomes a program DAG once it is assigned child nodes that satisfy all of the arguments required by the specification. Each child node must also be a valid program DAG. \textit{Constant}, \textit{Input}, and zero-arity \textit{Function} expressions are always valid program DAGs.

% For example, consider the following annotated Python function.

% \begin{verbatim}
%     def take(my_list: List, n: int) -> List:
%         return my_list[:n]
% \end{verbatim}

% The \texttt{my\_list} argument of the \texttt{take()} function can be of any \texttt{List} sub-type, including lists of any element type such as: \texttt{List[int]}, \texttt{List[str]}, \texttt{List[List[SomeClass]]}, etc.

% The \texttt{take()} function is wrapped in a \textit{Function} expression, $E_1$, and its type hint annotations are used as its specification. To build a program DAG out $E_1$ we assign a \textit{Constant} expression, $E_2$,  containing the list of integers $[3,2,1]$ with the specification of \texttt{List[int]} as the child corresponding to the \texttt{my\_list} argument. This is allowed because the return type of $E_2$ is a sub-type of the type required by the argument of $E_1$. We also supply an integer \textit{Input} expression to the $n$ argument of $E_1$.

% We can call this program DAG with different values for $n$ to get different slices of $[3,2,1]$. For example, when $n=1$, the DAG will return $[3]$ and when $n=2$, the DAG will return $[3,2]$. 

\subsection{Expression Reification}

Although a \textit{Function} expression has a specification for the arguments and return value, it is often possible to produce a more exact specification of the expression once the precise specification of one or more of its child expressions is known.

For example, the \texttt{Take} expression described in Figure~\ref{fig:first_three_expressions} is initially specified to return a \texttt{List}. However, it is known at development time that the \texttt{Take} method will return a list with the same element type as the list the method was called on. If a \texttt{MyList} expression from Figure~\ref{fig:first_three_expressions} is the child of a \texttt{Take} expression, the correct specification of the return type for the \texttt{Take} expression should be a \texttt{List[str]}.

\begin{figure}
    \centering
    \includegraphics[width=\columnwidth]{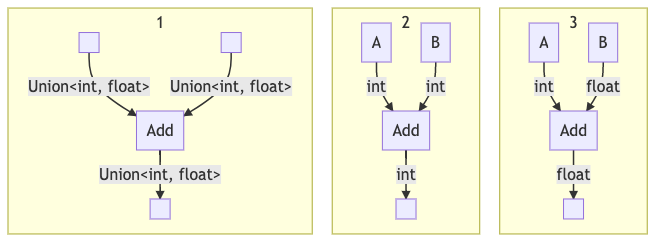}
    \caption{Box 1 diagrams an ``Add'' expression with its default specification and no arguments. Boxes 2 and 3 show two different possible reifications for the ``Add'' expression depending on the exact type of the arguments.}
    \label{fig:add_reification}
\end{figure}

We name this process of updating the specification of an expression instance ``expression reification'' because it concretizes attributes of the specification that become implicit given the additional information provided by the specification of the child nodes. Figure \ref{fig:add_reification} shows an example of different ways an arithmetic addition expression might be reified in a program DAG.

\begin{figure*}
    \centering
    \begin{tabular}{lll}
        Reification Rule & Description & Examples \\ \hline
        Pass Though & Expression returns the same type as a particular argument. & \texttt{abs, reverse, filter} \\
        Return Element & Expression returns the element type of a collection argument. & \texttt{first, last, nth, max, min} \\
        Args To Element & Expression argument must be the element type of the collection argument. & \texttt{index\_of, find, is\_in} \\
        Args To Same & Expression arguments must have the same type. & \texttt{concat, <, >}  \\
        List Of & Expression returns a list with an element type of another argument. & \texttt{list} \\
        Max Type & Expression returns the argument type that is highest in a hierarchy. & \texttt{+, -, *}\\
    \end{tabular}
    \caption{The set of reification rules used to in the prototype implementation of CBGP presented in this paper.}
    \label{fig:reification_rules}
\end{figure*}

Reification is implemented in CBGP as a set of zero or more ``reification rules'' that are assigned to each expression based on its logical behavior. Figure~\ref{fig:reification_rules} describes the reification rule types that are used in the CBGP implementation created for this research. Most function-like expressions either do not require reification, or require the exact reification described in one of the common rules. In other words, reification rules can be heavily reused throughout an application.

\subsection{The Push Compilation Process}
\label{sec:push}

Code Building Genetic Programming evolves a collection of expressions than can be ``compiled'' into a program DAG using a stack-based process inspired by the PushGP execution model~\cite{Spector:2005:push3}. PushGP uses this stack-based model to run the evolved programs, while CBGP uses this process to construct the program DAGs.

This compilation process translates a nested, sequential, collection of expressions into a type-safe, reified, program DAG that satisfies the overall specification of the program being evolved. This ensures that the program produced can be evaluated and is unlikely to produce run-time errors. 

The input to the DAG compilation process is a nested, sequential, structure of expressions and a specification of the return value for the desired program. Much like PushGP, the nested sequence of expressions is loaded onto the \texttt{exec} stack of a Push interpreter to be processed one element at a time. Unlike PushGP, the Push interpreter used by CBGP only contains 2 additional stacks: one for DAGs and one for anonymous function definitions.

While the \texttt{exec} stack is not empty, the top element is popped and processed depending on its type. \textit{Constant} and \textit{Input} instructions are pushed to the DAG stack. \textit{Function}, \textit{Method}, and \textit{Constructor} expressions undergo the following process:

\begin{enumerate}
    \item For each argument in the expression's specification:
    \begin{enumerate}
        \item The \texttt{DAG} stack is traversed.
        \item If the next DAG returns a sub-type of the expected argument's type, the DAG is a viable child to the expression and is removed from the \texttt{DAG} stack.
        \item The expression's specification is updated with its reification rules (if any).
        \item Step 1 is repeated until all arguments are satisfied by a child expression, or an argument is found that can't be satisfied by any DAG on the stack.
    \end{enumerate}
    \item If the set of child DAGs is incomplete:
    \begin{enumerate}
        \item The expression is discarded. 
        \item All stacks are reverted to their original states.
        \item Step 3 is skipped.
    \end{enumerate} 
    \item If all arguments are satisfied by a child node, 
    \begin{enumerate}
        \item The children are assigned to the expression, creating a valid DAG.
        \item  The new DAG is pushed to the \texttt{DAG} stack.
    \end{enumerate}  
\end{enumerate}

% \begin{figure}
%     \centering
%     \includegraphics[width=0.45\columnwidth]{img/push_compilation.png}
%     \caption{An example of the top items on the \texttt{DAG} stack before and after the processing of a multiplication expression from the top of the exec stack. }
%     \label{fig:push_compilation}
% \end{figure}

% Figure \ref{fig:push_compilation} depicts a single iteration of this process where the top two items of the \texttt{DAG} stack satisfy the arguments of the mutlip

If the top element of the \texttt{exec} stack is a list, it is pushed onto the \texttt{anonymous function} stack. It cannot be compiled into a program DAG because the required argument and return specifications are not known until the anonymous function is used by a specific higher order function.

When a \textit{Higher Order Function} expression is processed, it undergoes the following process:

\begin{enumerate}
    \item The DAG stack will be traversed to find a child expression that returns a collection type.
    \item If no collection type expression is found, the higher order function expression is discarded and the stacks are unchanged. Steps 3 through 5 are skipped.
    \item The \texttt{anonymous function} stack is traversed. For each list on the stack:
    \begin{enumerate}
        \item A new Push compilation process is made to compile the list into a DAG. This compilation process uses local input expressions as references to an element of the collection.
        \item If the nested compilation process produces a program DAG that returns the correct type\footnote{For example, \texttt{Filter} expressions must find a predicate anonymous function that returns a Boolean}, the DAG is assigned as the anonymous function body.
    \end{enumerate}
    \item If no viable anonymous functions are found on the anonymous function stack, the higher order function expression is discarded and any changes to stacks are reverted. 
    \item If the higher order function expression has both of the required child nodes, it becomes a valid DAG and is pushed to the DAG stack.
\end{enumerate}

If a local input expression is processed outside of the compilation of an anonymous function, it is ignored and has no effect on the stacks.

Once the exec stack is empty, the \texttt{DAG} stack will hold zero or more program DAGs. To find the single DAG that will be considered the program DAG, elements are popped from the \texttt{DAG} stack until a DAG is found whose return type is the same type (or a sub-type) of the return value for the desired program.

It is possible that this compilation does not produce a program DAG that satisfies the specification of the program being evolved. This situation must be handled in the error functions that are used to guide evolution toward a solution.

\section{Genetic Programming}

The novel capabilities of CBGP come from its expressive program DAG representation and unique process for safely building the DAGs as described in Section~\ref{sec:push}. The implementation of CBGP used for this research borrows all other aspects of the evolutionary computation from existing contemporary methods. The evolutionary population holds individuals defined by a linear genome. Error functions are used to assign errors to individuals. A parent selection strategy is used to pick individuals for variation. In summary, CBGP implements a typical generational evolutionary algorithm~\cite{Poli:2008:field-guide-to-gp}.

After evolution has found a solution individual, commonly used genome simplification methods are used to simplify individuals without sacrificing performance. Often the simplified genomes produce program DAGs with improved generalization performance~\cite{Helmuth:2017:simplification}.

\subsection{Genome Representation}

Individuals in CBGP use a derivative of the Plushy genome representation found in multiple PushGP systems~\cite{Pantridge:2018:plushi}. The genome is a flat sequence of expressions and structure tokens. The two kinds of structure tokens are \textit{OPEN} tokens and \textit{CLOSE} tokens. The Plushy genomes can be translated into a Push representation that can be compiled into a program DAG via the process outlined in Section~\ref{sec:push}.  

This linear structure allows for the use of a wide range of variation operators, and has been shown to yield better search results~\cite{plush, Helmuth:2018:PSU:3205455.3205603}. Also, the layers of indirection when translating Plushy genomes into Push code and compiling Push code into program DAGs ensure that type-safe computational graphs can be evolved without suffering from the bloat that accompanies using graph (or tree) structures directly.

\subsection{Evaluation of Program DAGs}

Once an individual's genome has been translated into Push code and compiled into a program DAG, it can be evaluated by an error function. The general structure of a typical error function is to evaluate the program on set of training cases containing input-output pairs. If the Push compilation process did not produce a program DAG that returns the required type for the given problem, the individual is assigned penalty errors for every training case. It is also possible for a program DAG to raise exceptions during run-time because not all expressions are defined for every value of their argument types. For example, a function with an integer argument might only return when given a natural integer and raise an exception when given a negative integer. If this happens during program DAG evaluation, the exception is caught and a penalty error is assigned for the training case.

One advantage CBGP has over PushGP is that the program DAGs only need to be compiled once and then the individual can be evaluated on all training cases. PushGP requires a separate execution of the stack-based interpreter for every training case. The computation graph representation of the programs produced during CBGP are much faster to compute than the stack-based execution in PushGP. Given that evaluation is the most expensive step of the evolutionary cycle, CBGP has the potential to dramatically reduce the cost of evolution for program synthesis tasks.

\section{Producing Source Code}

One appealing aspect of G3P systems is their ability to produce source code that is potentially readable by humans and is portable between codebases. This capability implies there is a possibility that a sufficiently capable GP system could eventually be used in cooperation with human developers to contribute to the same codebase. 

The program DAGs synthesized by Code Building Genetic Programming are roughly analogous to abstract syntax trees. Using knowledge of the host-language's syntax, it is possible to produce source code from a program DAG. This process makes the output of CBGP human interpret-able and portable, similar to G3P.

As mentioned previously, program DAGs evolved by CBGP contain expressions that wrap existing functions and methods. Thus, the source code representation of a program DAG includes calls to these functions and methods. This further motivates the potential cooperation between human programmers and a program synthesis framework as mentioned in Section~\ref{sec:reflection}.

\section{Benchmarks}
\label{sec:benchmarks}

For every benchmark problem, the CBGP system was configured to use the following evolutionary settings. 

\begin{figure}[h]
    \vspace*{-1.0em}
    \centering
    \begin{tabular}{ll}
        Parameter & Setting \\ \hline
        Runs & 31 \\
        Generations & 300 \\
        Population Size & 1000 \\
        Selection & Lexicase \\
        Variation & UMAD \\
        Random Training Cases & 100 \\
        Test Cases & 1000 \\
    \end{tabular}
    \vspace*{-1.0em}
\end{figure}

With the exception of the relatively low number of runs per problem, these settings are comparable to those used by other GP frameworks in previous publications. We leave the study and calibration of these settings for CBGP as future work.

We tested a prototype implementation of CBGP on two small sets of benchmark problems. The first is a novel set of benchmarks which are designed to demonstrate CBGP's ability to handle polymorphic types and integrate with existing codebases. The second is a sample of problems taken from the ``General Program Synthesis Benchmark Suite'' which has become a standard benchmark set for program synthesis GP systems since its introduction in 2015~\cite{Helmuth:2015:BenchmarkSuite}.

The novel benchmarks presented in this section are similar to typical utility functions that might be implemented in real world applications as part of a larger system. The ``General Program Synthesis Benchmark Suite'' problems are derivatives of introductory academic assignments~\cite{Helmuth:2015:BenchmarkSuite}. In the following subsections we will describe the novel benchmarks created for CBGP.

\begin{figure}
    \centering
    \begin{tabular}{lll}
         DateTime & TimeDelta & Path \\ \hline
         year() -> int & days() -> int & to\_str() -> str \\
         month() -> int & seconds() -> int & abspath() -> Path \\
         day() -> int & & split() -> List[str] \\
         hour() -> int & & basename() -> str \\
         minute() -> int & & dirname() -> Path \\
         second() -> int &  & isabs() -> bool \\
         & & join(other: Path) -> Path \\
    \end{tabular}
    \caption{APIs of the classes defined for the benchmark problems. These classes are annotated versions of existing classes provided by the Python standard library. This collection of classes are meant to demonstrate CBGP's ability to interface with pre-existing codebases.}
    \label{fig:methods}
\end{figure}

\subsection{Days Between}

The ``Days Between'' problem is designed to demonstrate CBGP's ability to work with pre-defined types and classes. The problem prompt is as follows:

\begin{quote}
    Given two DateTime objects, return the absolute number of days between them.
\end{quote}

The API of \texttt{DateTime} class and the related \texttt{TimeDelta} class can be found in Figure~\ref{fig:methods}. If should also be noted that \texttt{DateTime} and \texttt{TimeDelta} objects can be compared with comparison functions (ie. $<$, $\leq$, $>$, $\geq$, $==$) and shifted with arithmetic functions (ie. \texttt{add}, \texttt{sub}). 

To our knowledge, no program synthesis GP methods have shown a capability to work with date and time data types, without explicit configuration and extension.

\subsection{Filter Bounds}

The ``Filter Bounds'' problem shows CBGP's ability to produce functions that work with different instances of a polymorphic type. The ``Filter Bounds'' problem also requires use of higher order functions. The problem prompt is as follows:

\begin{quote}
    Given a list of elements that are all of the same comparable type, $T$, and two instance of type $T$ representing a lower and upper bound, filter the list to the elements that fall between two bounds (inclusively). For example, if given the list \verb|[6,5,4,3,2,1]|, the lower bound \verb|3|, and the upper bound \verb|5| the result should be \verb|[5,4,3]|. Also, given the list \verb|["a","b","c"]|, the lower bound \verb|"x"|, and the upper bound \verb|"zzz"| the result should be an empty list.
\end{quote}

The datasets of training and test cases use a variety of comparable types for $T$. The evolved solution is expected to work generically for all lists of comparable elements.

\subsection{Prefix Paths}
R
The ``Prefix Paths'' problem is designed with similar goals to the ``Days Between'' problem, except with the added complexity of requiring the use of a higher order function. The problem prompt is as follows:

\begin{quote}
    Given a \texttt{Path} object representing a root directory and a list of file names (as strings), return a list of \texttt{Path} objects that join the root path and each filename. The resulting list of \texttt{Path} objects should be in the same order as the given filenames. For example, given a root of \verb|Path("/tmp")| and a list of files 
    
    \verb|["log.txt", "data.csv"]|, the result should be 
    
    \verb|[Path("/tmp/log.txt"), Path("/tmp/data.csv")]|.
\end{quote}

The API for the `Path` class used in this benchmark can be found in Figure \ref{fig:methods}. The solution program should work with absolute and relative root paths.

\section{Results}

We present solution rates of a Code Building Genetic Programming system on the three novel benchmarks in the following table.

\begin{figure}[h]
    \vspace*{-1.0em}
    \centering
    \begin{tabular}{lll}
        Problem & Solutions & Rate \\ \hline
        Days Between & 31/31 & 100\% \\
        Filter Bounds & 31/31 &  100\% \\
        Prefix Paths & 30/31 & 96.8\%
    \end{tabular}
    \vspace*{-1.0em}
\end{figure}

These benchmark problems are not complex and these high solution rates don't necessarily indicate a strong overall search performance of CBGP. Instead, these problems are meant to demonstrate particular novel capabilities of CBGP described in Section~\ref{sec:benchmarks}. To highlight the successful demonstration of these capabilities, Figure~\ref{fig:code_output} contains generated Python code which was produced using a solution DAG from one evolutionary run and simple string formatting rules that transcribe a program DAG into valid source code. We see usages of functions, methods, constants, constructors, and higher order functions in the generated source code.

\begin{figure*}
    \centering
    \begin{verbatim}
def days_between(dt1, dt2):
    return abs(sub(dt1, dt2).days())
    
def prefix_files(root, filenames):
    return map(lambda _0: root.join(Path(_0)), filenames)    
    
def filter_bounds(lst, lower, upper):
    return filter(lambda _0: lt(lt(_0, lower), le(_0, upper)), lst)
    
def replace_space_with_newline(input1):
    return sub(len(input1), print_tap(input1.replace(" ", "\n", -87)).count("\n")) 

def negative_to_zero(input1):
    return map(lambda _0: max(bool2int(not(float2bool(0.5724738469524758))), _0), input1)
    \end{verbatim}
    \caption{A small sample of solution code snippets produced by CBGP on the benchmark problems described in Section~\ref{sec:benchmarks}. The functions that are not built-in Python functions (ie. \texttt{bool2int}, \texttt{print\_tap}) are simple wrapper functions that add the necessary annotations. Notice that CBGP generated the name ``\_0'' for the arguments to lambda functions.}
    \label{fig:code_output}
\end{figure*}

\begin{figure}
    \centering
    \includegraphics[width=\columnwidth]{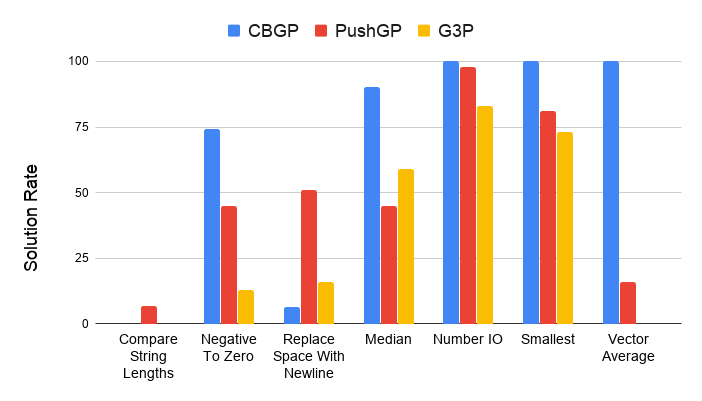}
    \caption{Solution rates between CBGP, PushGP, and G3P on a subset of the General Program Synthesis Benchmark Suite.}
    \label{fig:ss_solutions}
\end{figure}

We also tested CBGP on seven problems from the ``General Program Synthesis Benchmark Suite'' in order to compare performance between CBGP and other contemporary program synthesis methods. Figure~\ref{fig:ss_solutions} shows the solution rates between CBGP, PushGP, and G3P.

The solution rates for CBGP are significantly higher (with a P-value of 0.05) than PushGP for \textit{Negative To Zero}, \textit{Median}, \textit{Smallest}, and \textit{Vector Average}. CBGP performs significantly worse than PushGP on \textit{Compare String Lengths} and \textit{Replace Space With Newline}.

Code Building GP is not significantly worse that G3P on any of the problems we experimented with and finds significantly more solutions for \textit{Negative To Zero}, \textit{Median}, \textit{Number IO}, \textit{Smallest}, and \textit{Vector Average}.

It should be noted that differences in supported functions/instructions make comparison between CBGP, PushGP, and G3P problematic. The CBGP implementation created for this research uses a subset of Python's built-in functions and types to create expressions, in line with the design goals of the system\footnote{A listing of python functions used in our CBGP prototype can be found in the source code files at https://github.com/erp12/CodeBuildingGeneticProgramming-ProtoType/tree/master/push4/library}. PushGP and G3P use manually curated instruction sets and grammars that are designed not to ``cheat'' at solving the benchmarks with instructions that are too close to a solution. This practice is useful when comparing methods, but is unrepresentative of real-world applications, for which any augmentation or restriction of the instruction set would be acceptable if it produces results that meet a current need.

In most cases, the improved performance of CBGP can easily be explained by the differences in supported functions. For example, Python has a built-in \texttt{sum} function that will sum a list of numbers. This function is made available to CBGP but neither PushGP, nor G3P, choose to support an equivalent operation. This gives CBGP a significant advantage on the ``Vector Average'' problem. 

Another crucial difference between PushGP, G3P, and CBGP is configuration of which operations to use for a particular problem. Both PushGP and G3P select a subset of supported operations and data types to include before an evolutionary run. This configuration varies for each benchmark problem. The CBGP experiments presented in this paper use the entire set of core functions and classes that were defined and annotated to be identical to the built in capabilities of the Python language. This puts CBGP at a disadvantage because its search space has not been narrowed to the relevant operatoins.

We suggest that future research should test CBGP more rigorously, including some experimentation with a set of expressions that matches what is available in PushGP and G3P for each problem.

\section{Discussion and Future Work}

The theoretical capabilities of Code Building Genetic Programming promise a wider range of potential applications than any other program synthesis GP system. The initial indication from our empirical results show that CBGP can demonstrate these applications, at least for simple problems. To our knowledge, CBGP is the first inductive programming method, genetic programming or otherwise, to demonstrate the ability to synthesize programs that utilize arbitrary preexisting data types and an ability to handle polymorphism.

Furthermore, the ability for CBGP to gather its own set of supported expressions by leveraging technologies like reflection greatly reduces the requirements on external configuration. This makes CBGP a much simpler program synthesis framework, potentially suitable for real world applications by non-expert practitioners.

With regard to problem solving capabilities, our initial crude results indicate that CBGP is comparable or superior on some problems, while severely lacking on others. More rigorous experiments on a wider set of benchmarks is required to understand the general advantages and disadvantages between CBGP and its contemporaries. This research is ongoing.

The following subsections will discuss areas of future research which would improve the overall quality of a Code Building Genetic Programming system.

\subsection{Post-processing of Program DAGs}

Computational graphs, like the ones constructed by CBGP, are common representations of executable procedures. One advantage to using computational graphs is the ability to canonicalize and optimize the graphs for improved performance.

Big data frameworks, like Spark, utilize lazy evaluation of execution plans consisting of map-reduce operations to allow for query optimization~\cite{Zaharia:2012:RDD}. The optimizer mutates the DAG into a more efficient computation that has the same behavior. This optimization can also be used to canonicalize an execution plan so that caching strategies can detect if a equivalent computation has already been performed.

Code Building GP could eventually utilize similar optimization algorithms to produce programs that are more resource efficient and easier to convert into source code. The code in Figure~\ref{fig:code_output} has multiple instances of unnecessary function calls. The presented solution to the ``Negative to Zero'' problem has an 4 node sub-DAG that will always return a zero. Genome simplification was unable to address these issues, but it is likely that a more sophisticated canonicalization process would.

\subsection{More Expressive Specifications}

The prototype CBGP implementation uses data type based specifications. Types are a relatively weak form of specification because some functions are not defined across the entire argument space. As specification tooling matures, it will be beneficial to implement CBGP such that it utilizes more information than data types. This will allow evolution to avoid most, if not all, run-time errors in its program DAGs which will reduce the use of penalty errors and smooth out the search space.

\section{Conclusion}

In this paper, we present Code Building Genetic Programming and show early demonstrations of its unique capabilities. The benchmarks clearly demonstrate the use of preexisting functions and classes in evolved program DAGs, as well as the transcription of DAGs into valid, type-safe, Python code. This would not be possible without the introduction of expression reification and the stack-based compilation process.

Although our comparison to other program synthesis frameworks is flawed, the problem solving ability of CBGP is promising. Regardless, we recognize the value of first-class specifications and tools like reflection in making program synthesis methods more adaptive to a given use case. We hope these lessons will help propel the field beyond the academic benchmarks towards real-world applications.

Finally, we direct future research towards rigorous evaluation of CBGP, DAG canonicalization, richer specification, and attempted collaborations between a CBGP system and a human programmer on a complex, real-world application.

\begin{acks}
We thank Bill Tozier, Nicholas McPhee, and the members of the Hampshire College Institute for Computational Intelligence for discussions that advanced this work. This material is based upon work supported by the National Science Foundation under Grant No. 1617087. Any opinions, findings, and conclusions or recommendations expressed in this publication are those of the authors and do not necessarily reflect the views of the National Science Foundation.
\end{acks}

% \newpage
\bibliographystyle{ACM-Reference-Format}
\bibliography{cbgp-bib} 

\end{document}